\documentclass[review]{elsarticle}

\journal{}

\usepackage{subfigure}
\usepackage{amsmath}
\usepackage{multirow}
\usepackage{color}

\begin{document}

\begin{frontmatter}

\title{A Probability Distribution and Location-aware ResNet Approach for QoS Prediction}


\author[mymainaddress]{Wenyan Zhang}

\author[mymainaddress]{Ling Xu\corref{mycorrespondingauthor}}
\cortext[mycorrespondingauthor]{Corresponding author}
\ead{xuling@cqu.edu.cn}
\author[mymainaddress]{Meng Yan}

\author[mymainaddress]{Ziliang Wang}

\author[mymainaddress]{Chunlei Fu}

\address[mymainaddress]{School of Big Data \& Software Engineering, ChongQing University, \par Chongqing, 400044, China}

\begin{abstract}
    In recent years, the number of online services has grown rapidly, invoke the required services through the cloud platform has become the primary trend. How to help users choose and recommend high-quality services among huge amounts of unused services has become a hot issue in research. Among the existing QoS prediction methods, the collaborative filtering(CF) method can only learn low-dimensional linear characteristics, and its effect is limited by sparse data. Although existing deep learning methods could capture high-dimensional nonlinear features better, most of them only use the single feature of identity, and the problem of network deepening gradient disappearance is serious, so the effect of QoS prediction is unsatisfactory. To address these problems, we propose an advanced probability distribution and location-aware ResNet approach for QoS Prediction(PLRes). This approach considers the historical invocations probability distribution and location characteristics of users and services, and first use the ResNet in QoS prediction to reuses the features, which alleviates the problems of gradient disappearance and model degradation. A series of experiments are conducted on a real-world web service dataset WS-DREAM. The results indicate that PLRes model is effective for QoS prediction and at the density of 5\%-30\%, which means the data is sparse, it significantly outperforms a state-of-the-art approach LDCF by 12.35\%-15.37\% in terms of MAE.
\end{abstract}


\end{frontmatter}


\section{Introduction}\label{sec:Introduction}
\noindent
\par With the rise of various cloud application platforms, the number of various services increases rapidly. At the same time, users are more likely to invoke the services of these cloud platforms to implement relevant functions instead of downloading various applications. However, there are many candidate services in the cloud environment, which makes it difficult for users to choose a suitable service. So researchers are trying to find some ways to help users find better services among many with the same functionality.

\par Quality of service (QoS) is the non-functional evaluation standard of service, including service availability, response time, throughput, etc. Its value is often affected by the network environment of the user and the service. In different network environments, QoS values generated by different users may vary greatly even if the invoked service is the same one. Therefore, it is meaningful to predict QoS values of candidate services before the user invokes a service, which can help the target user distinguish the most suitable service among different functionally equivalent services according to the predicted QoS results\cite{8640087, Wang2019, DBLP:journals/amc/DingXCZY14, DBLP:journals/monet/YinCXWZM20,8691437,DBLP:journals/eswa/XuYDXH16}. At present, QoS value has become a pivotal criterion for service selection and service recommendation, and QoS prediction has also been applied in plenty of service recommendation systems.

\par In recent years, collaborative filtering(CF) methods are widely used for QoS prediction\cite{DBLP:conf/icsoc/ZouJNWPG18, DBLP:journals/tpds/ZhuHZL17,DBLP:conf/icsoc/WuXCCZ17,DBLP:journals/access/TangLYX19, DBLP:journals/sensors/CaiDX19,DBLP:journals/www/LiuSLLXZZS19,DBLP:journals/nrhm/SunWWMH16}, which relies on the characteristics of similar users or items for target prediction. In QoS prediction, the collaborative filtering methods match similar users or services for target users or services first, and then uses the historical invocations of these similar users or services to calculate the missing QoS. Because of its strong pertinence to the target user and item, CF is often used in personalized recommendation systems. However, CF can only learn low-dimensional linear features, and its performance is usually poor in the case of sparse data. To address these problems, several QoS prediction approaches based on deep learning have been proposed, and these approaches have been proved to be very effective in QoS prediction\cite{DBLP:journals/monet/YinCXWZM20,DBLP:conf/ijcai/XueDZHC17,zhang2019location, xiong2018deep}. Yin et.al\cite{DBLP:journals/monet/YinCXWZM20} combined Matrix Factorization(MF) and CNN to learn the deep latent features of neighbor users and services. Zhang et.al\cite{zhang2019location} used multilayer-perceptron(MLP) capture the nonlinear and high-dimensional characteristics. Although the existing deep learning methods have improved in QoS prediction, they will not perform better when the network is deep due to the inherent gradient disappearance of deep learning. Inspired by the deep residual learning(ResNet)\cite{DBLP:conf/cvpr/HeZRS16}, which is widely used in the field of image recognition, we realize that the reuse feature can effectively alleviate the gradient disappearance problem in deep learning. ResNet consists of multiple residual blocks, each of which contains multiple shortcuts. These shortcuts connect two convolution layers to realize feature reuse, prevent the weakening of original features of data during training, and achieve the purpose of alleviating gradient descent.

Among the existing deep learning approaches, most of them\cite{DBLP:journals/monet/YinCXWZM20, DBLP:conf/ijcai/XueDZHC17} only use ID as the characters, and a few methods\cite{zhang2019location} introduce the location information. However, users and services in the same region often have similar network status, which provides a crucial bases for finding similar neighborhoods. Therefore, the introduction of geographic position is often helpful for achieving higher accuracy in QoS prediction. In addition, none of these methods consider using probability distribution as the characteristic. Probability distribution refers to the probability of QoS predictive value in each interval, which is calculated by the historical invocations of the target. For example, if a user's invocation history indicates that the response time is almost always less than $0.5s$, the probability of missing value less than $0.5s$ is much higher than the probability of missing value greater than 0.5s. Therefore, the introduction of probability distribution could reflect the historical invocation of users and services. For QoS prediction, historical invocation is the most important reference basis, so it is necessary to introduce probability distribution as a feature in QoS prediction.

Therefore, in this paper, we propose a probability distribution and location-aware ResNet approach(PLRes) to better QoS prediction. 
First, PLRes obtains the information of target users and services, including identifier information, geographical location and historical invocation, and calculates the probability distribution of target users and services according to the historical invocation. Then PLRes embedded ID and location characteristics into high-dimensional space, and concatenated the embedded feature vectors and probability distribution vectors. Next, the ResNet is used to learn the nonlinear feature of the combined characteristics. Finally, PLRes is exploited to predict the missing QoS value.


The contributions of this paper are as follows:
\begin{itemize}
    \item We calculate the probability distribution of target users and services and take them as the characteristics of QoS prediction. This characteristic reflects the network conditions of target users and services, and reduces the error of Qos prediction.
    \item We propose a novel probability distribution and location-aware QoS prediction approach PLRes, which is based on ResNet. In our approach, we use the identifier, location information and probability distribution as the characteristics, and first introduce the ResNet for QoS prediction, which uses the idea of feature reuse to enhance the features in the process of model training. This enables our model to learn nonlinear high-dimensional characteristics well and get better results when the network depth increases.
    \item We validated the PLRes on a real-world dataset, WS-DREAM\footnote{http://wsdream.github.io/}, and compared the predictive performance with various existing classical QoS prediction methods and the state-of-the-art deep learning approach LDCF\cite{zhang2019location}. Experimental results show that our method outperforms the state-of-the-art approach for QoS prediction significantly.
\end{itemize}

\par The remainder of the paper is organized as follows. In Section \ref{sec:ProposedModel}, we describe our QoS prediction model in detail. In Section \ref{sec:Experiments}, we introduce the experiment setup, followed by experimental results and discussion in Section \ref{sec:Results}. In Section \ref{sec:Discussion}, we discussed the reasons why our model works. In Section \ref{sec:RelatedWork}, we provide an overview of related works. In the last section, we conclude our work and provide an outlook on directions for future work.

\section{Proposed Approach}\label{sec:ProposedModel}
\noindent
\par In this section, we give a detailed description of the proposed approach.

\subsection{The Framework of the Proposed Model}
\begin{figure*}[htbp]
    \centerline{\includegraphics[width=1.0\textwidth]{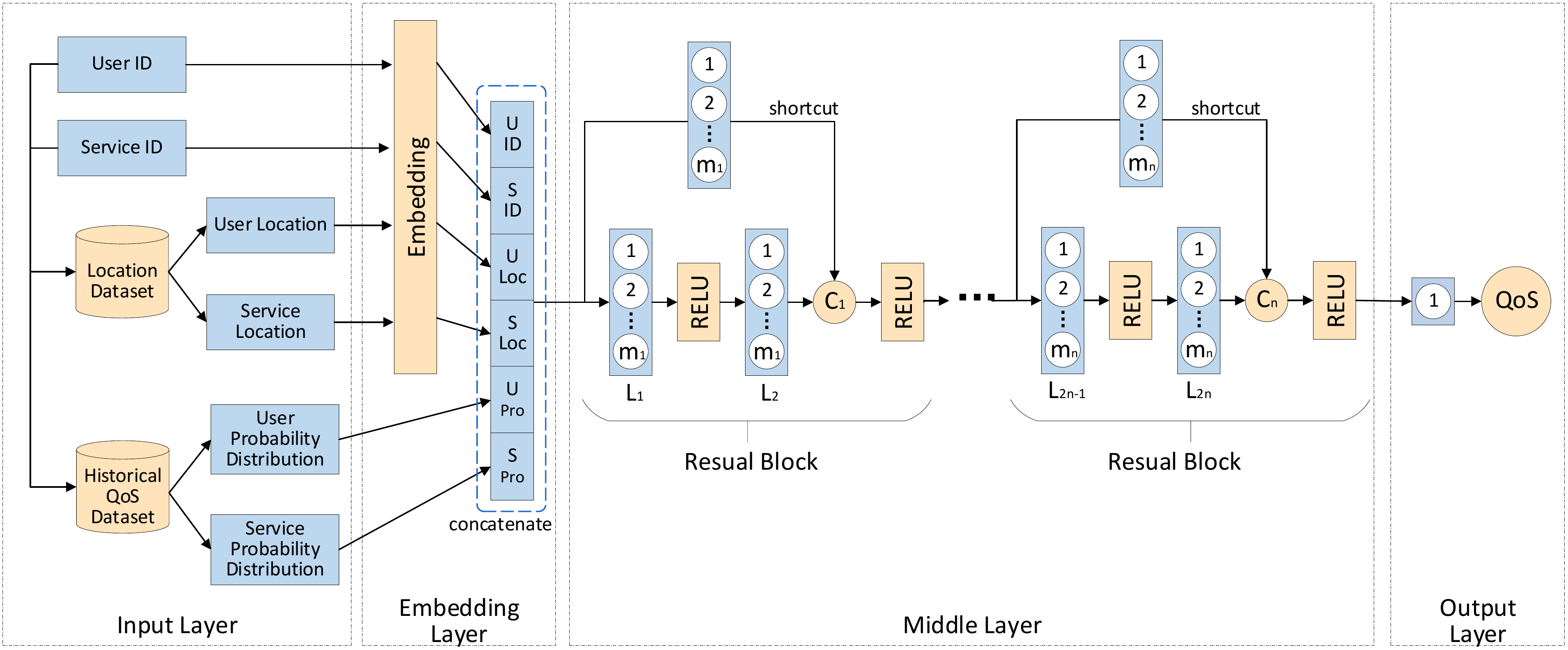}}
    \caption{The framework of the proposed model.}
    \label{framework}
\end{figure*}

The overall architecture of PLRes is shown in Figure \ref{framework}, which includes the input layer, the embedding layer, the middle layer, and the output layer.

The process of PLRes can be expressed as: the model receives a pair of user and service characteristics(including ID, location and probability distribution) as input, then embedded the identity and location features in the high-dimensional space respectively. Next, the embedded vectors and the probability distribution are concatenated into a one-dimensional vector. PLRes learns the one-dimensional feature and finally, gives the prediction result according to the learned characteristic rule. 
The following subsections describe the model details. Section \ref{Input_Layer} and \ref{Embedding_Layer} describe the input and embedding of features respectively. Section \ref{Middle_Layer} describes the learning process of the model. Section \ref{Output_Layer} describes the final prediction and output details, and Section \ref{Model_Learning} describes the setting of the model optimizer and loss function. 


\subsection{Input Layer}
\label{Input_Layer}
\par The input layer is primarily responsible for receiving features. The features we selected include the user ID, the user's location, the user's probability distribution, the service ID, the service's location and the service's probability distribution. Both the user ID and the service ID are represented by an assigned integer. So only one neuron is needed for the input of both. The location information of the user and the service is represented by country and AS(Autonomous System), so the location information of the user and the service each needs two neurons. Probability distribution needs to be calculated based on historical invocations. The calculation is described in Section \ref{sec:preprocessing}, and the number of neurons it required in the input layer is related to the number of QoS value intervals $K$.

\subsection{Embedding Layer}
\label{Embedding_Layer}
\par The embedding layer mainly does two jobs: embedding ID and location features into the high-dimensional space, and feature fusion for all features.
At first, it maps the discrete features into high-dimensional vectors. There is no doubt that in our dataset, ID, country and AS are all discrete features, which need to be encoded to be the data that deep network computing can be used.

\par In the embedding layer, we use one-hot to encode these four features(the ID and location of the user and the service) and then embed them into high-dimensional space. One-hot is one of the most common methods to encode discrete features, which makes the calculation of the distance between feature vectors more reasonable. In one-hot encoding, each value of the characteristic corresponds to a bit in a one-dimensional vector, only the position whose value corresponding to the current characteristic is 1, and the rest are set to 0. We use $u$, $s$, $u_l$ and $s_l$ to represent the one-hot coded user identify, service identify, user location and service location respectively. In the embedding process, the random weights are generated first, and the weights are adjusted continuously according to the relationship between features in the model learning process, and the features are mapped into high-dimensional dense vectors. The embedding process could be shown as follows:
\begin{equation}
    I_{u} = f_e(W_{u}^{T}u+b_u)
    \label{embedding1}
\end{equation}
\begin{equation}
    I_{s} = f_e(W_{s}^{T}s+b_s)
    \label{embedding3}
\end{equation}
\begin{equation}
    L_{u} = f_e(W_{u_l}^{T}u_l+b_{u_l})
    \label{embedding2}
\end{equation}
\begin{equation}
    L_{s} = f_e(W_{s_l}^{T}s_l+b_{s_l})
    \label{embedding4}
\end{equation}
where $I_{u}$, $I_{s}$ represents the identify embedding vector of user and service, and $L_{u}$, $L_{s}$ is the location embedding vector of user and service respectively. $f_e$ represents the activation function of embedding layer; $W_u$, $W_{u_l}$, $W_s$ and $W_{s_l}$ represents the embedding weight matrix; $b_u$, $b_{u_l}$, $b_s$ and $b_{s_l}$ represents the bias term.

Then the model uses the concatenation mode to fuse the features into a one-dimensional vector and passed to the middle layer. In addition to the ID and location characteristics embedded in the high-dimensional space described above, the probability distribution characteristics of users and services are also included. We use $P_{u}$ and $P_{s}$ to represent the probability distributions of users and services. The concatenated could be expressed as:
\begin{equation}
    x_0 = \Phi(I_{u},I_{s},L_{u},L_{s},P_{u},P_{s}) = \begin{bmatrix} I_{u} \\ I_{s} \\ L_{u} \\ L_{s} \\ P_{u} \\ P_{s}\end{bmatrix}\quad
    \label{concatenate}
\end{equation}

\subsection{Middle Layer}
\label{Middle_Layer}
\par The middle layer is used to capture the nonlinear relationship of features, and we used ResNet here. ResNet is mainly used for image recognition and uses a large number of convolutional layers.
In image recognition, the characteristics are composed of neatly arranged pixel values, while the feature we use is a one-dimensional vector, which is not suitable for convolutional layer processing, so we only use the full connection layer. 

Our middle layer is composed of multiple residual blocks, as shown in Figure \ref{framework}, each of which consists of a main road and a shortcut. In the main road, there are two full connection layers and two 'relu' activation functions; The shortcut contains a full connection layer. Before the vector in the main path passes through the second activation function, the original vector is added to the main path vector by the shortcut, which is the process of feature reuse. 

In a residual block, the number of neurons in the two fully connected layers is equal. Since the number of neurons in two vectors must be the same to add, when the original feature takes a shortcut, a full connection layer is used to map it so that it can be successfully added to the vector of the main path. For the $i$th residual block, the full connection layers in the main road are the $(2i-1)$th layer and $(2i)$th layer of the middle layer. We used $m_i$ to represent the number of neurons in the full connection layer and $C_i$ to represent the sum of vectors in the $i$th residual block.
\begin{equation}
    M_i = W_i^Tf_i(W_i^Tx_{i-1}+b_i)+b_i,\; i=1,2,\dots,n
    \label{m_i}
\end{equation}
\begin{equation}
    S_i = W_i^Tx+b_i,\; i=1,2,\dots,n
    \label{s_i}
\end{equation}
\begin{equation}
    C_i = M_i+S_i,\; i=1,2,\dots,n
    \label{c_i}
\end{equation}
\begin{equation}
    x_i = f_i(C_i),\; i=1,2,3,\dots,n
    \label{x_i}
\end{equation}
where $M_i$ and $S_i$ respectively represents the vector of the main path and shortcut before adding the vectors in the $i$th residual block; $C_i$ represents the sum of two vectors of the $i$th residual block; $x_i$ represents the output of the $i$th residual block, and $x_0$ represents the output of the embedding layer; $f_i$ represents the activation function of the $i$th residual block, and $W_i$ and $b_i$ represents the corresponding weight matrix and bias term.



\subsection{Output Layer}
\label{Output_Layer}
The output layer of our model has only one neuron to output the final result. The output layer is fully connected to the output of the last residual block in the middle layer. In this layer, we use the linear activation function. The equation is:
\begin{equation}
    \hat{Q}_{u,s} = W_o^Tx_n+b_o
    \label{result}
\end{equation}
where $\hat{Q}_{u,s}$ denotes the predictive QoS value of the service invoked by the user; $x_n$ represents the output of the last residual block in the middle layer; $W_o$ and $b_o$ denote the weight matrix and bias term of the output layer. 

\subsection{Model Learning}
\label{Model_Learning}
\subsubsection{Loss Function Selection}
Since the prediction of QoS in this paper is a regression problem, we choose the loss function from MAE and MSE according to the commonly used regression loss function. Their formulas are expressed as Eq.~\ref{MAE} and Eq.~\ref{MSE}. The difference between the two is the sensitivity to outliers, and MSE will assign a higher weight to outliers. In QoS prediction, outliers are often caused by network instability, and sensitivity to outliers tends to lead to overfitting, which affects the accuracy of prediction. Therefore, we choose MAE as the loss function, which is relatively insensitive to abnormal data. We will also discuss the effect of the two in Section \ref{subsec:Evaluation_Metrics} and Section \ref{subsubsec:Loss_Function}.

\subsubsection{Optimizer Selection}
Common optimizers include SGD, RMSprop, Adam\cite{DBLP:journals/corr/KingmaB14}, etc. We used the Adam optimizer in our proposed model. As an adaptive method, Adam optimizer works well for sparse data. Compared with SGD, Adam is faster. And compared with RMSprop, Adam performs better with the bias-correction and momentum when the gradient is sparse.

\section{Experimental Setup}\label{sec:Experiments}
\noindent
This section presents four investigated research questions(RQs), the experimental dataset, the compared baseline models, and the widely used evaluation measures.

\subsection{Research Questions}
\noindent\textbf{RQ1.~How effective is our proposed PLRes?}

The focus of the first RQ is the effect of our model for QoS prediction. If PLRes shows advantages over traditional QoS prediction models and the state-of-the-art QoS predict model LDCF, it is proved that the learning by PLRes is beneficial for QoS prediction.

\noindent\textbf{RQ2.~How does the probability distribution affect the accuracy of prediction?}

This RQ aims to evaluate if the introduction of probability distribution contributes to a better performance. To analyze the impact of the probability distribution, we run the PLRes without this characteristic and compare the predicted results to the previous results to determine whether the performance has declined.

\noindent\textbf{RQ3.~How does the location affect the accuracy of prediction?}

This research focuses on the impact of location characteristics for QoS prediction. We set up a model with geographical position information removed, which only use ID and probability distribution as features for training. The test results of this model are compared with those of PLRes model to judge whether location information contributes to the improvement of QoS prediction model performance.

\noindent\textbf{RQ4.~How does the reuse of characteristics affect the accuracy of prediction?}

The way to reuse characteristics in the proposed PLRes model is to introduce shortcuts to the traditional Deep Neural Networks(DNN). RQ4 investigates whether the introduction of shortcuts contributes to improve the model performance. If the PLRes(uses shortcuts) is better than the results of traditional DNN(without shortcuts), it proves that characteristic reuse improves the model.

\noindent\textbf{RQ5.~How do different parameter settings affect the model effectiveness?}

The proposed PLRes contains three important parameters: the depths, loss function and learning rate. RQ5 aims to investigating the impact of different parameter settings and providing a better choice for each parameter.

\subsection{Experimental Dataset}
\subsubsection{Original Dataset}
\begin{figure}[htbp]
    \centerline{\includegraphics[width=0.3\textwidth]{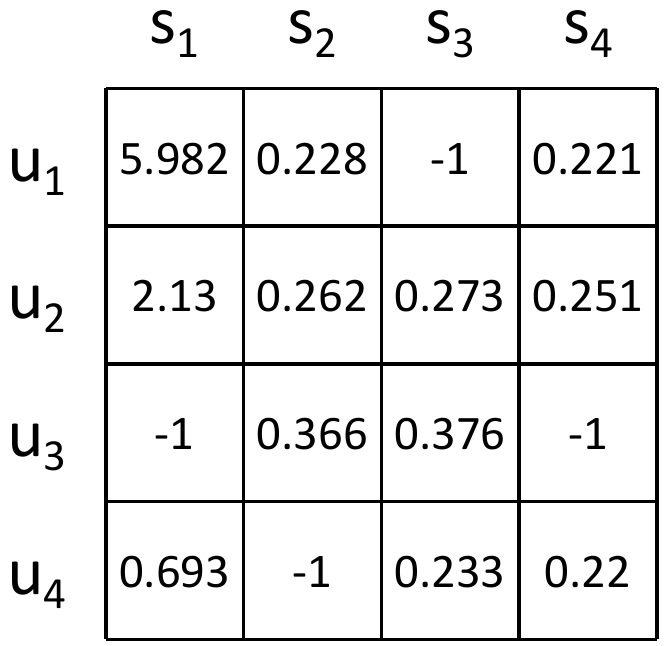}}
    \caption{The respond time matrix.}
    \label{fig:rtMatrix}
\end{figure}

We used the WS-DREAM dataset, which is the QoS dataset of real-world Web services collected by Zheng et al\cite{DBLP:journals/tsc/ZhengMLK11}. The dataset contains 1,873,838 available QoS (including response time and throughput) values of Web services collected from 339 users on 5825 services. In our experiments, we used the response time to verify our method. Take response time as an example, the form of the QoS matrix is shown in the Figure \ref{fig:rtMatrix}, -1 represents the invalid response time, which means the user did not invoke the service or the response time timeout for invoking the service. For user $u_1$ in Figure \ref{fig:rtMatrix}, user $u_1$ has not effectively invoked service $s_3$, while the response time of invoking service $s_1$, $s_2$ and $s_4$ is $5.982$, $0.228$ and $0.221$, respectively. 

\begin{figure}[htbp]
    \centering
        \subfigure[Information of users]{
            \includegraphics[width=1.0\textwidth]{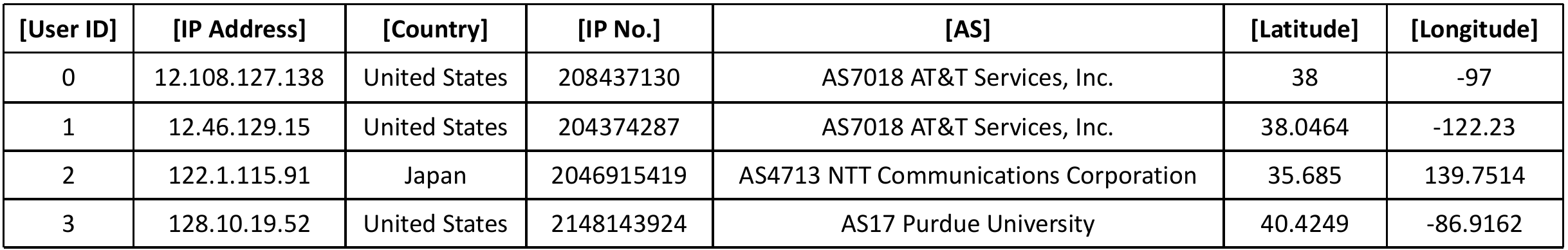}}
        \subfigure[Information of Services]{
            \includegraphics[width=1.0\textwidth]{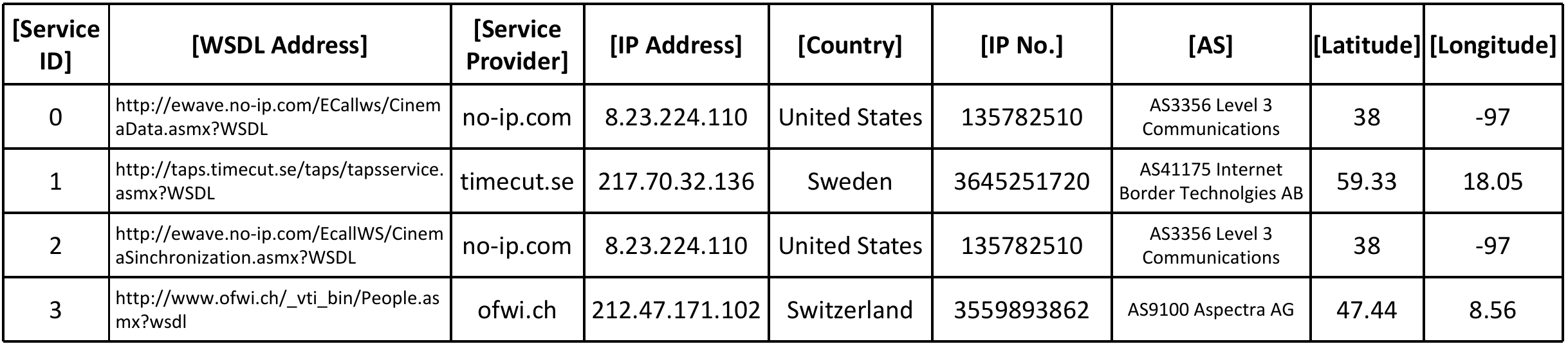}}
    \caption{Information of users and services.}
    \label{fig:Information}
\end{figure}

The dataset also includes other information about users and services. The user information and service information are shown in Figure \ref{fig:Information}. The user information includes [ID, IP Address, Country, IP NO., AS, Latitude, Longitude] and the service information includes [ID, WSDL Address, Service Provider, IP Address, Country, IP NO., AS, Latitude, Longitude]. We use [Country, AS] as the location characteristics of the user and service. 

\subsubsection{Preprocessing}\label{sec:preprocessing}

\begin{figure}[htbp]
    \centerline{\includegraphics[width=0.9\textwidth]{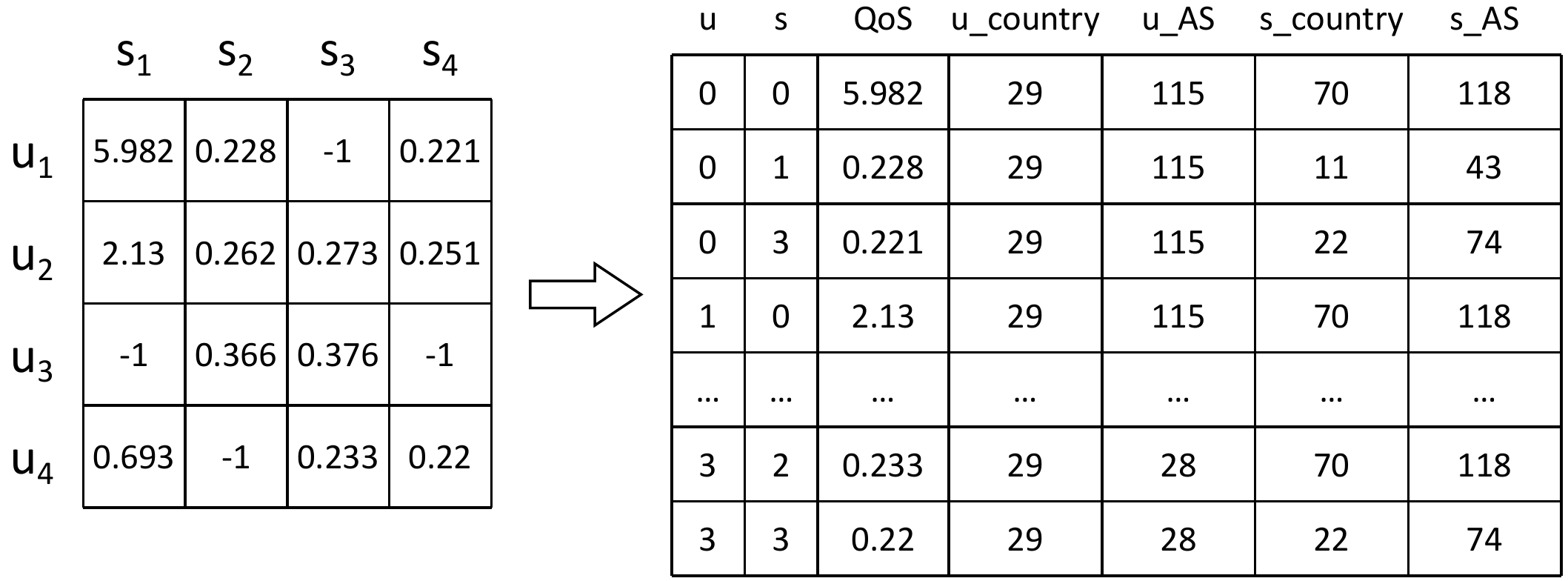}}
    \caption{Transformed the matrix to invoked records.}
    \label{datatrans}
\end{figure}

In the data preprocessing, we merge the required information from the original data(including the original QoS data set, user information and service information) and converted them into the form of the invocation record. As shown in Figure \ref{datatrans}, the final invocation record converted from QoS matrix is represented as [user ID, service ID, QoS value, user Location, service Location], and locations include country and AS. All IDs and locations in the dataset are assigned unique numbers.

In addition, we need to calculate and store the probability distribution of each user and service. We take the historical QoS distribution of target user and service as the QoS probability distribution in the prediction. In the experiment, the training set of the model is used as historical invocations. Before calculating the QoS distribution, the number of intervals K should be confirmed first, and then the range of each interval could be confirmed according to K. When calculating the probability distribution of the target user or service, it is necessary to obtain all the historical invocations of the target and count the number of each interval.
The QoS distribution of the target can be obtained by calculating the proportion of each interval in the total number of invocations. The probability calculation method can be defined as follows:
\begin{equation}
P_{u}(k)=\frac{Num(Q_{u},k-1,k)}{Num(Q_{u},0,K)}\label{Probability}
\end{equation}
\begin{equation}
P_{s}(k)=\frac{Num(Q_{s},k-1,k)}{Num(Q_{s},0,K)}\label{Probability}
\end{equation}
where $P_{u}(k)$, $P_{s}(k)$ denotes the probability of the Qos appearing in $k$th interval according to the historical invocations of user $u$ and service $s$; $Num(Q_{u},k-1,k)$, $Num(Q_{s},k-1,k)$ denotes the number of the Qos appearing in $k$th interval according to the historical invocations of user $u$ and service $s$; $Num(Q_{u},0,K)$, $Num(Q_{s},0,K)$ denotes the total number of the user $u$'s invocations and the service $s$'s respectively. Take the first user in Figure \ref{user_pro}, $user258$ as an example, the QoS used by this distribution is response time, with K set to 10. Since the dataset used records the maximum response time as 20s, it is set to an interval every 2 seconds. The number of invocation records by $user258$ in each interval is [5366, 92, 22, 5, 8, 15, 33, 4, 2, 2], and the total number is 5549. So the $user258$ in 10 interval probability is [96.7\%, 1.66\%, 0.4\%, 0.09\%, 0.14\%, 0.27\%, 0.59\%, 0.07\%, 0.04\%, 0.04\%].

\subsection{Comparison Methods}
\par We select the following QoS prediction methods to compare their performance with our method:
\begin{itemize}
    \item \textbf{UIPCC(User-Based and Item-Based CF)\cite{DBLP:journals/tsc/ZhengMLK11}:} This approach is a classic collaborative filtering, which computes similar users and similar services by PCC, and combines them to recommend services to target users. It is the combination of UPCC(User-Based CF) and IPCC(Item-Based FC).

    \item \textbf{PMF(Probabilistic Matrix Factorization)\cite{DBLP:conf/nips/SalakhutdinovM07}:} This is a very popular method of recommending fields. MF is to factor the QoS matrix into an implicit user matrix and an implicit service matrix, and PMF is to introduce the probability factor into MF.
    \item \textbf{LACF\cite{DBLP:conf/icws/TangJLL12}:} This is a location-aware collaborative filtering method. The difference of the method and traditional collaborative filtering is that it uses the users close to the target user on the geographic location as similar users, and the services close to the target service on the geographic location as similar services.
    \item \textbf{NCF\cite{DBLP:conf/www/HeLZNHC17}:} This method combines CF and MLP, inputs implicit vectors of users and services into MLP, and uses MLP to learn the interaction between potential features of users and services.
    \item \textbf{LDCF\cite{zhang2019location}:} This is a location-aware approach that combines collaborative filtering with deep learning. It is a state-of-the-art QoS prediction method, and we take it as our baseline model.
 \end{itemize}

\par Among these approaches, UIPCC and PMF are content-based and model-based collaborative filtering methods, respectively, LACF and LDCF are location-aware methods, and NCF and LDCF are neural network-related models.

\subsection{Evaluation Metrics}\label{subsec:Evaluation_Metrics}
\par The prediction of QoS can be classified as a regression problem, so we use the Mean Absolute Error(MAE) and Root Mean Squared Error (RMSE) to measure the performance of the prediction. MAE and RMSE are defined as:
\begin{equation}
    MAE=\frac{1}{N}{\Sigma_{u,s}|{Q_{u,s}}-\hat{Q}_{u,s}|}\label{MAE}
\end{equation}
\begin{equation}
    MSE=\frac{1}{N}{\Sigma_{u,s}({Q_{u,s}}-\hat{Q}_{u,s})^2}\label{MSE}
\end{equation}
\begin{equation}
    RMSE=\sqrt{MSE}=\sqrt{\frac{1}{N}{\Sigma_{u,s}({Q_{u,s}}-\hat{Q}_{u,s})^2}}\label{RMSE}
\end{equation}
where $Q_{u,s}$ is the actual QoS value of service $s$ observed by user $u$, $\hat{Q}_{u,s}$ is the predictive QoS value of service $s$ observed by user $u$, and $N$ denotes the total number of QoS.

\section{Experimental Reasults}\label{sec:Results}
\noindent
In this section, a series of experiments are designed to answer the four questions raised in Section \ref{sec:Experiments}, and the experimental results will be presented and analyzed.

\subsection{RQ1: Model Effectiveness}
\par In the experiments, we use the same data to train the models of comparison methods and PLRes, and test them with the same testing set. For the CF methods need to find similar users or services(UIPCC and LACF), the number of neighbours is set to 10. And for the deep learning method(NCF and LDCF), we set the number of MLP layers as six and the number of neurons in each layer as [256,128,128,64,64,1]. For the MF method(PMF and NCF), we set the number of implicit features as 10. For the parameters that all models need to be used, we set the learning rate to be 0.001, the batch size to be 256 and the maximum number of iterations to be 50. As for the loss function and optimizer, we use the default parameters for each model to ensure that they work well.

\begin{table}[htbp]
    \centering
    \caption{Experimental results of different Qos prediction approach}
    \subtable[MAE]{
    \setlength{\tabcolsep}{4mm}{
    \begin{tabular}{c|c|c|c|c|c|c}
    \hline
    \textbf{density}&\textbf{5\%}&\textbf{10\%}&\textbf{15\%}&\textbf{20\%}&\textbf{25\%}&\textbf{30\%} \\
    \hline
    \hline
    UIPCC   &   0.625   &   0.581   &   0.501   &  0.450    &   0.427   &   0.411\\
    PMF     &   0.570   &   0.490   &   0.460   &   0.442   &   0.433   &   0.428\\
    LACF    &   0.630   &   0.560   &   0.510   &   0.477   &   0.456   &   0.440\\
    NCF     &   0.440   &   0.403   &   0.385   &   0.359   &   0.344   &   0.338\\
    LDCF    &   0.406   &   0.371   &   0.346   &   0.336   &   0.325   &   0.314\\
    \textbf{PLRes}&\textbf{0.356}&\textbf{0.317}&\textbf{0.297}&\textbf{0.285}&\textbf{0.279}&\textbf{0.273}\\   
    \hline 
    \end{tabular}}}
    \subtable[RMSE]{
    \setlength{\tabcolsep}{4mm}{
    \begin{tabular}{c|c|c|c|c|c|c}
    \hline
    \textbf{density}&\textbf{5\%}&\textbf{10\%}&\textbf{15\%}&\textbf{20\%}&\textbf{25\%}&\textbf{30\%} \\
    \hline
    \hline
    UIPCC   &   1.388   &   1.330   &   1.250   &   1.197   &   1.166   &   1.145\\
    PMF     &   1.537   &   1.320   &   1.230   &   1.179   &   1.156   &   1.138\\
    LACF    &   1.439   &   1.338   &   1.269   &   1.222   &   1.188   &   1.163\\
    NCF     &   1.333   &   1.274   &   1.242   &   1.218   &   1.184   &   1.177\\
    LDCF    &   1.297   &   1.223   &   1.184   &   1.164   &   1.132   &   1.113\\
    \textbf{PLRes}&\textbf{1.244}&\textbf{1.187}&\textbf{1.140}&\textbf{1.123}&\textbf{1.108}&\textbf{1.094}\\   
    \hline 
    \end{tabular}}
    }
    \label{table:compare}
\end{table}

\begin{figure}[htbp]
    \centering
        \subfigure[MAE]{
            \includegraphics[width=0.45\textwidth]{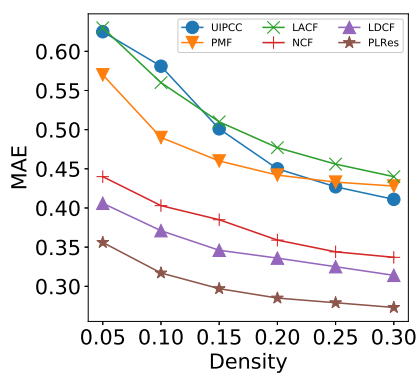}
            \label{compare_mae}}
        \quad
        \subfigure[RMSE]{
            \includegraphics[width=0.45\textwidth]{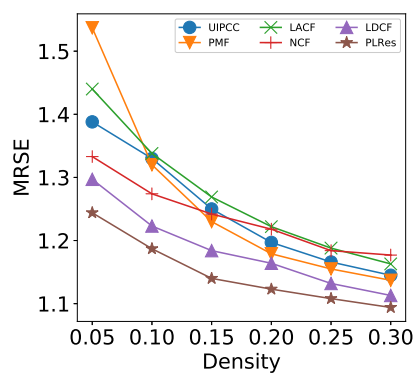}
            \label{compare_rmse}}
    \caption{Performance comparison of 6 methods.}
    \label{fig:compare}
\end{figure}
    
Table \ref{table:compare} shows the detailed test results of the above approaches and our model in six different densities. Figure \ref{fig:compare} show the advantages of our method more intuitively. According to the comparison result, with the increase of density and the training data, the MAE and RMSE performance of these methods are all improved, and PLRes always performs best at the same density. 

Can be observed in the Figure \ref{compare_mae}, the performance comparison of MAE, the models using deep learning(NCF, LDCF and PLRes) are all below 0.45 at the density of 5\%, which perform better than the other three models(UIPCC, PMF, LACF), whose MAE were all above 0.55. Similarly, at other densities, the models using deep learning are more effective. This strongly proves the ability of deep learning to fit nonlinear features in QoS prediction.

In terms of the performance comparison of RMSE, it can be observed from the Figure \ref{compare_rmse} that the performance of deep learning models are better than those of CF models at the density of 5\% and 10\%. It reflects that the CF method is difficult to perform well under sparse density, while the deep learning method greatly alleviates this problem. When the density is greater than 10\%, although the CF models gradually outperform the deep learning method NCF, LDCF and PLRes still perform best. This may be related to the introduction of location characteristics and probability distribution characteristics.

It is worth mentioning that compared with the baseline model LDCF, PLRes improves MAE performance by 12.35\%, 14.66\%, 14.17\%, 15.37\%, 14.24\% and 13.22\%, RMSE performance by 4.10\%, 2.95\%, 3.24\%, 3.48\%, 2.13\% and 1.78\% respectively under the density of 5\%-30\%. Furthermore, we apply the Wilcoxon signed-rank\cite{wilcoxon1992individual} test on the prediction results of PLRes and LDCF at the density of 5\%(the QoS matrix is extremely sparse) to analyze the statistical difference between the two models. The p-value is less than 0.01, which indicates that the improvement of PLRes against LDCF is statistically significant.

\subsection{RQ2: Effect of Probability Distribution}
\par In order to examine the impact of probability distribution, we removed the characteristics of probability distribution, and only took user ID, service ID, user location, service location as the factors to conducted the experiment with the same ResNet model. In the two experiments, we set the same training data for training, set the maximum number of iterations to 50, and saved the model with the best testing results. 
\par
\begin{figure}[htbp]
    \centering
        \subfigure[MAE]{
            \includegraphics[width=0.45\textwidth]{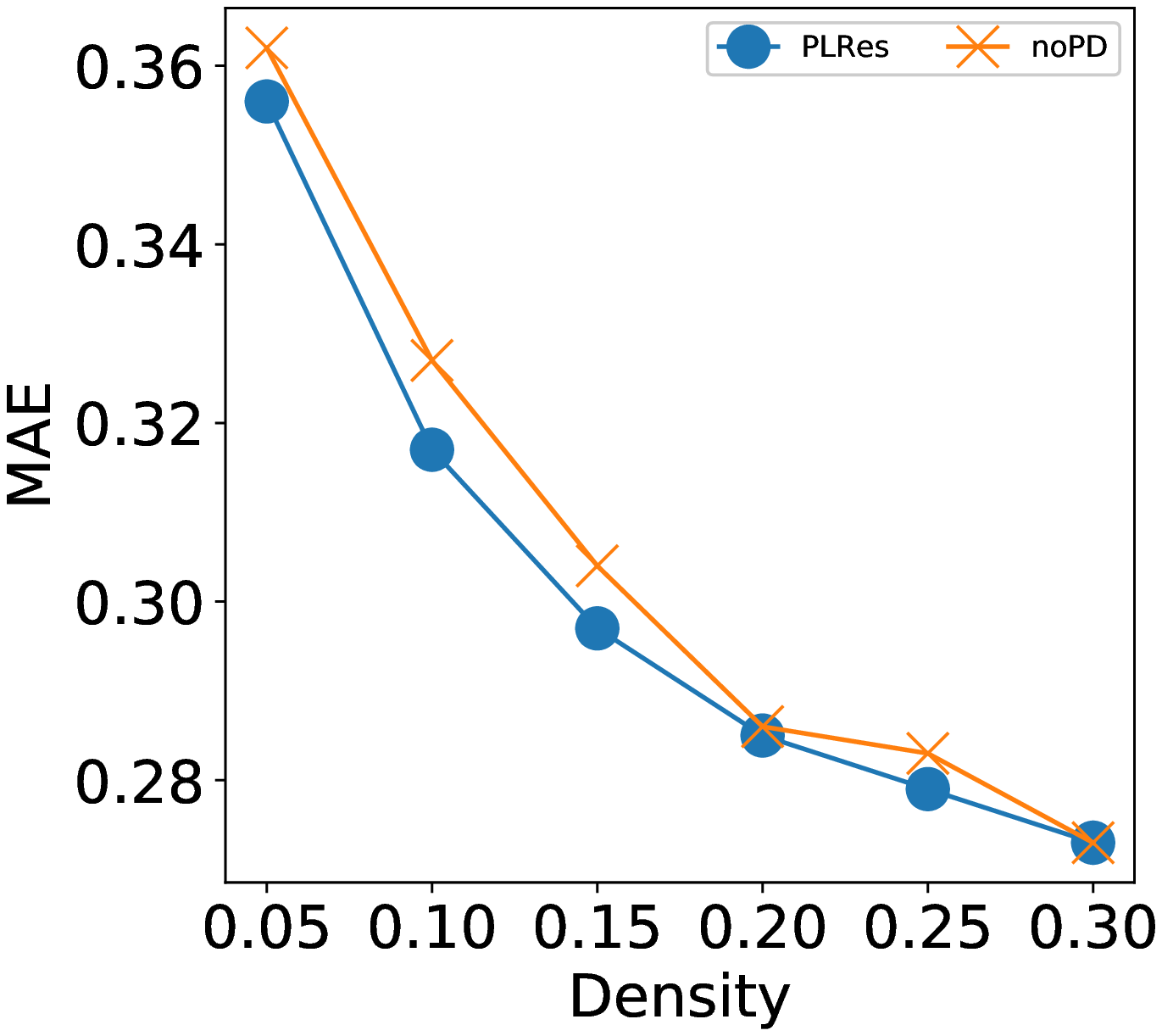}
            \label{pro_mae}}
        \quad
        \subfigure[RMSE]{
            \includegraphics[width=0.45\textwidth]{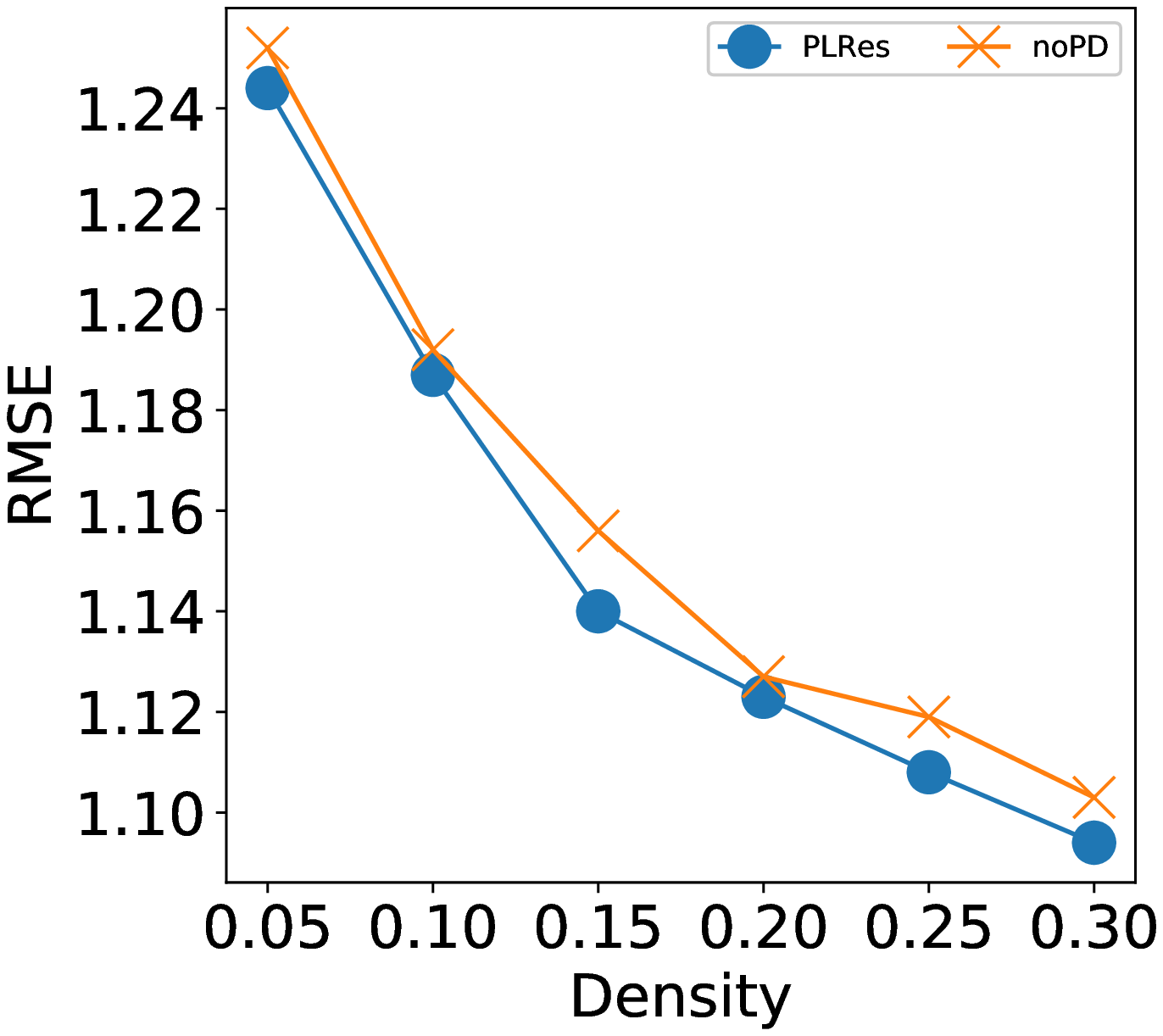}
            \label{pro_rmse}}
    \caption{Impact of probability distribution.}
    \label{fig:Pro}
\end{figure}

 The results of the two experiments at different densities are shown in the Figure \ref{fig:Pro}. From the results, the performance of the model with the probability distribution as the feature has better performance than the model without the probability distribution feature at all six densities. The results fully prove that the introduction of probability distribution is beneficial to improve the performance of the model.

\subsection{RQ3: Effect of Location}\label{subsec:effect_loc}
\par
\begin{figure}[htbp]
    \centering
        \subfigure[MAE]{
            \includegraphics[width=0.45\textwidth]{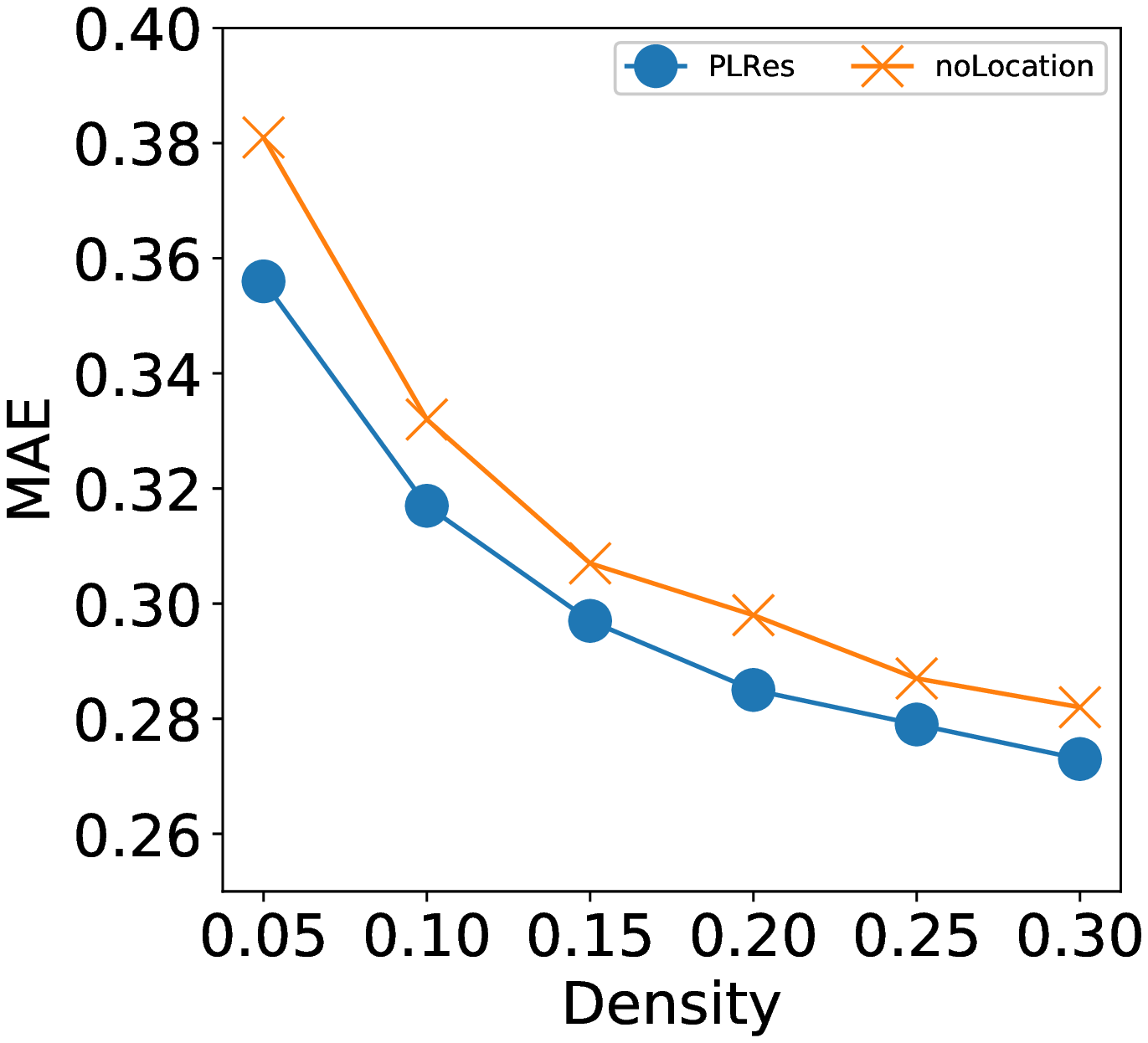}
            \label{loc_mae}}
        \quad
        \subfigure[RMSE]{
            \includegraphics[width=0.45\textwidth]{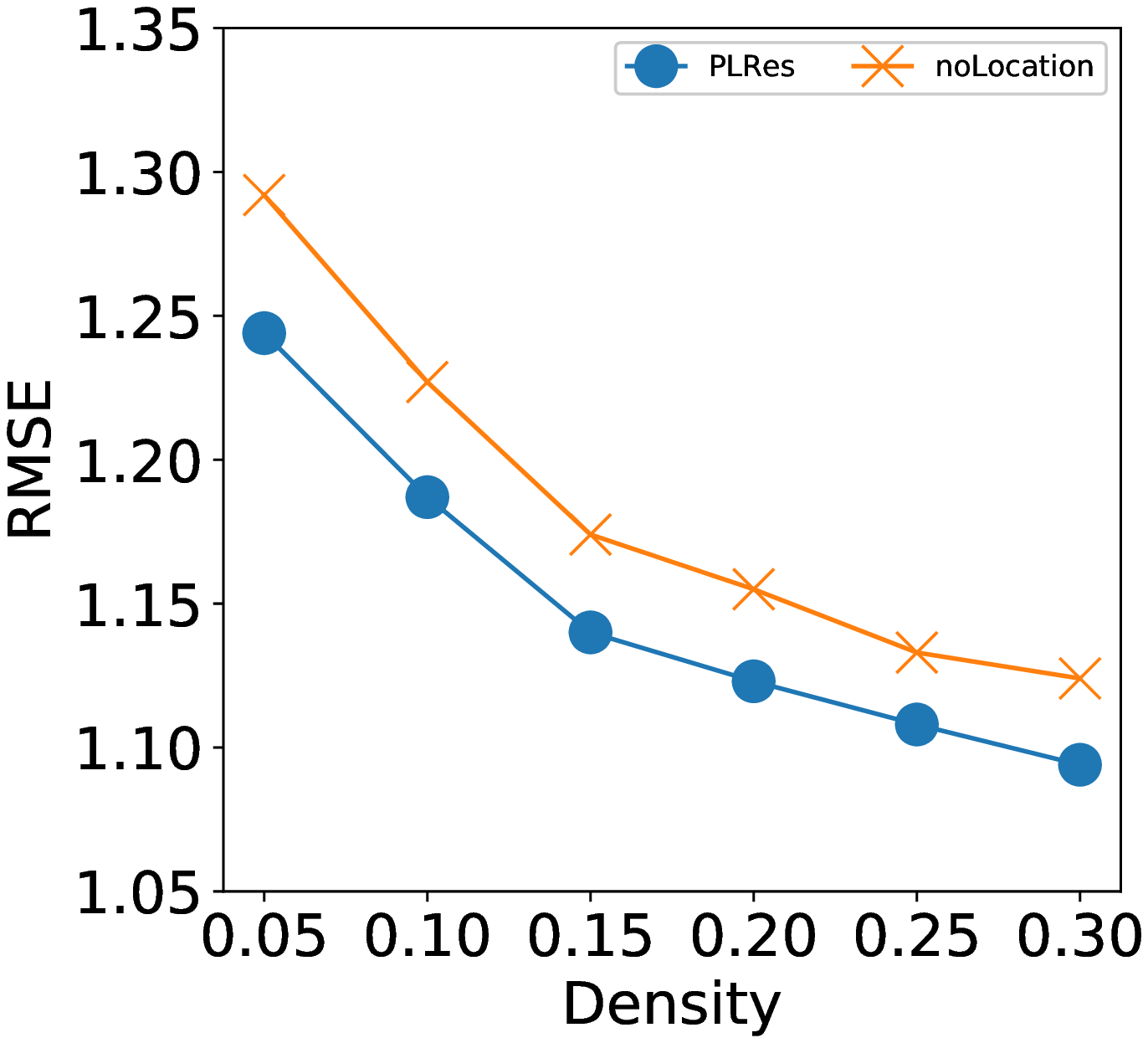}
            \label{loc_rmse}}
    \caption{Impact of location.}
    \label{fig:loc}
\end{figure}

 We will verify the importance of location information to our model in this section. We try to train the model using only ID and probability distribution as characteristics, and compare the testing results with PLRes. Figure \ref{fig:loc} shows our test results, in which the test results of PLRes are represented by blue lines, and the test results of the model that only used ID and probabiliity distribution as the characteristics are represented by orange lines. Figure \ref{loc_mae} is the MAE result and Figure \ref{loc_rmse} is the RMSE result. At 5\%-30\% of the density we tested, although the trend of performance change with density increase is similar, the MAE and RMSE  performance of the model decreased significantly when we take out the location information, indicating that location characteristics do have an improved effect on QoS prediction. 

\subsection{RQ4: Effect of Shortcuts}



\begin{figure}[htbp]
    \centering
        \subfigure[MAE]{
            \includegraphics[width=0.45\textwidth]{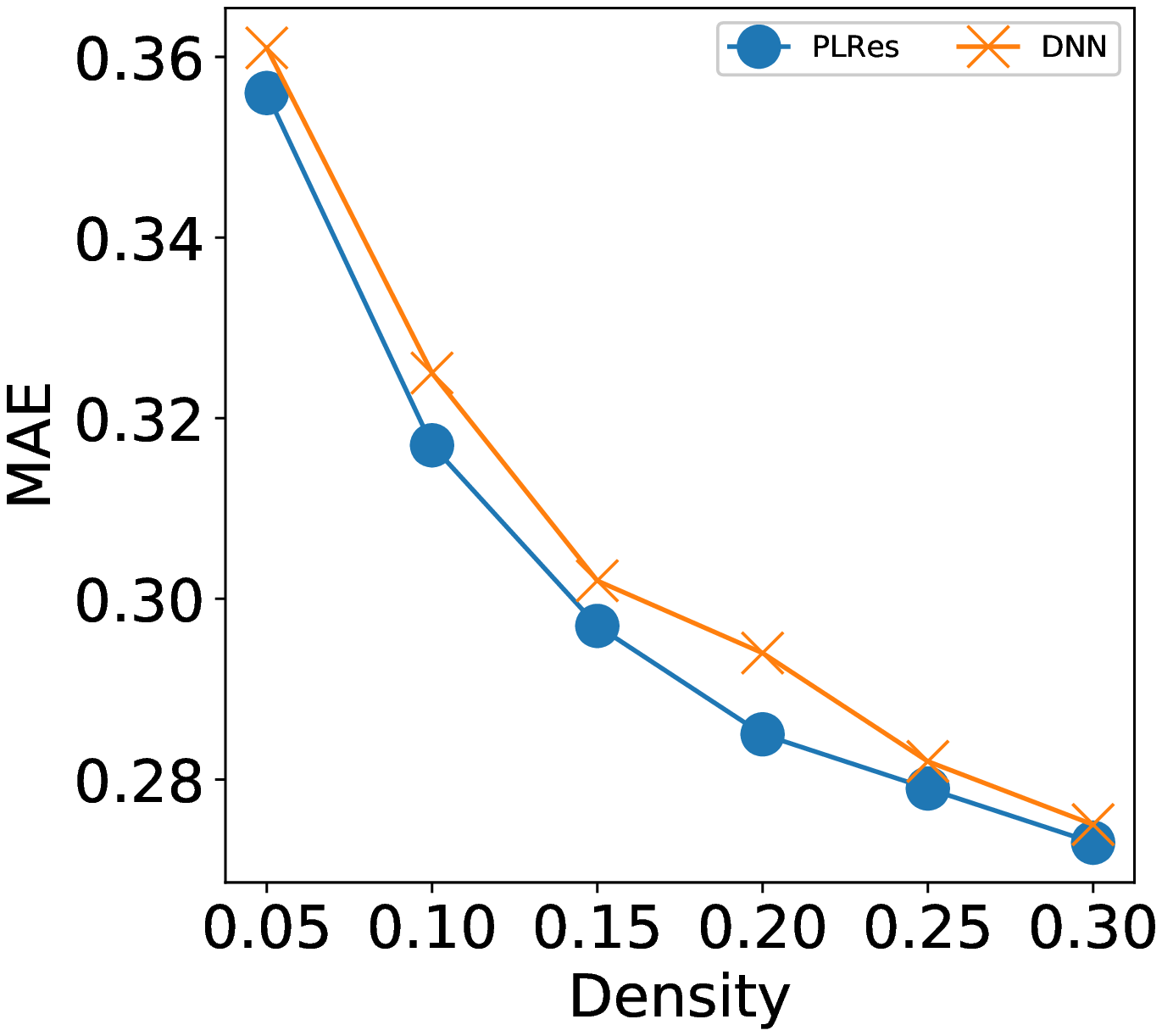}}
        \quad
        \subfigure[RMSE]{
            \includegraphics[width=0.45\textwidth]{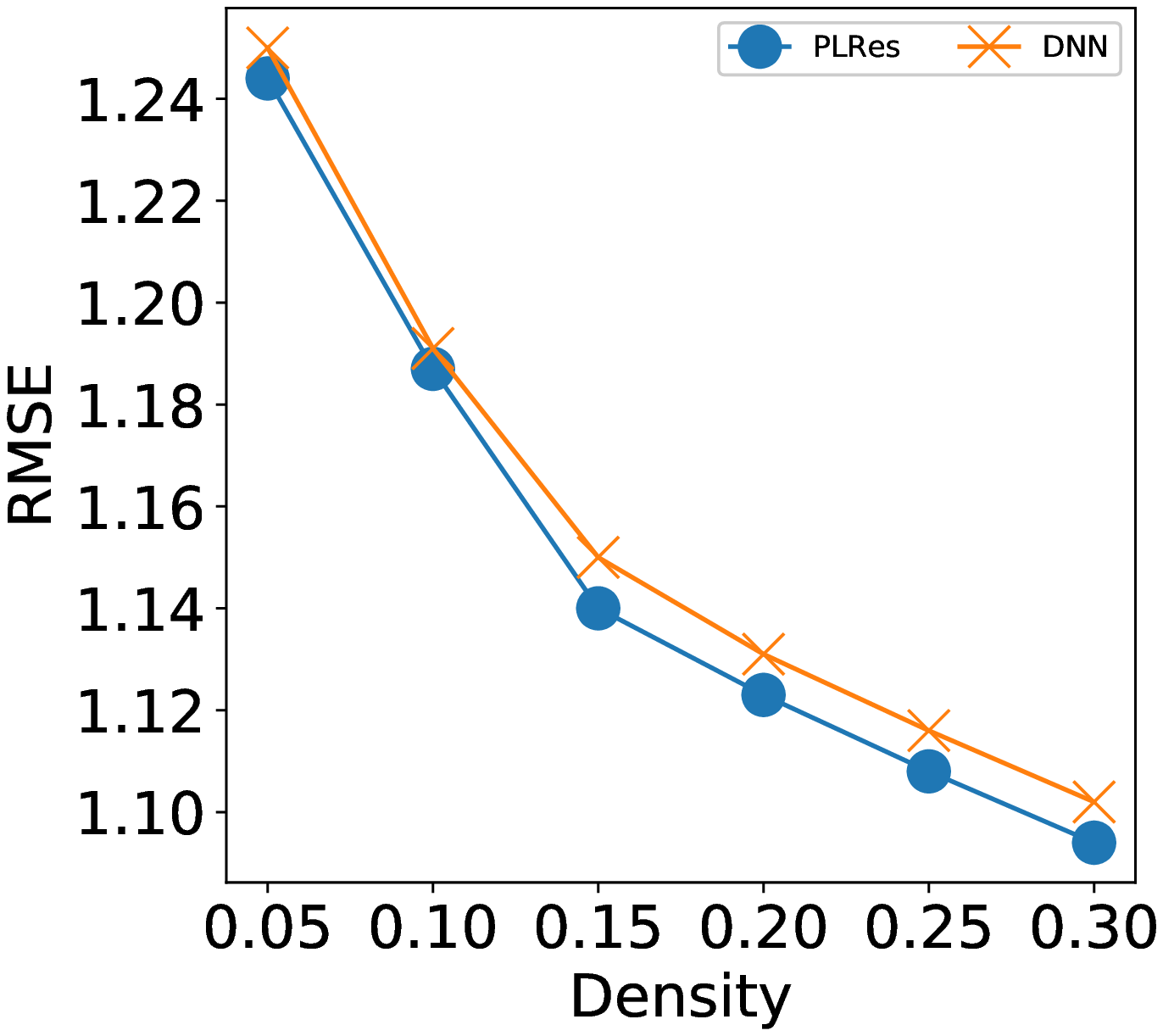}}
    \caption{Impact of shortcuts.}
    \label{fig:shortcuts}
\end{figure}

The method of feature reuse in ResNet is to use shortcuts, which add original features directly to trained data. In this section, we discuss the impact of shortcuts on our experimental results. In this set of experiments, we used the DNN and the ResNet to learn the same dataset respectively, so as to prove the effectiveness of the shortcuts. We set the PLRes to use two residual blocks, each of which contains two full connection layers, so in the DNN we set the number of hidden layers to 4. In PLRes, the number of neurons in the two residual blocks is [128,64], and the number of neurons in each hidden layer in the DNN is [128, 128, 64, 64]. The testing results are shown in Figure \ref{fig:shortcuts}.

Although shortcuts are used to improve the performance of our network in deeper networks, increasing the number of network layers also means increasing the cost of time and space. Therefore, we hope that the introduction of shortcuts can also help improve the performance of the model even when the network is shallow. In this experiment, only 2 residual blocks(4 hidden layers) are used, which also significantly improves the performance, indicating that feature reuse is effective in this model, and the introduction of shortcut improves the performance of the model. Furthermore, we discuss the impact of shortcuts on the deepening of network layers in Section \ref{sec:depths}.

\subsection{RQ5: Impact of Parameter Settings}
\subsubsection{Impact of Depths}\label{sec:depths}
\par
\begin{figure}[htbp]
    \centering
        \subfigure[MAE]{
            \includegraphics[width=0.45\textwidth]{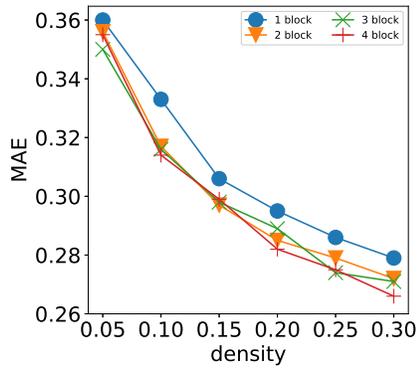}
            \label{pro_mae}}
        \quad
        \subfigure[RMSE]{
            \includegraphics[width=0.45\textwidth]{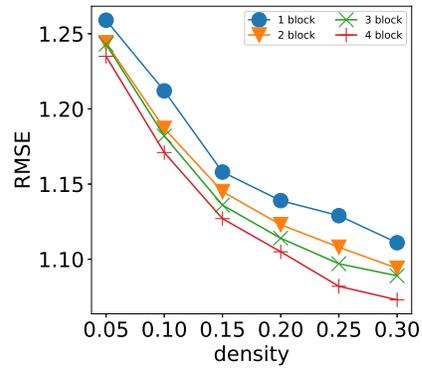}
            \label{pro_rmse}}
    \caption{Impact of the number of residual blocks.}
    \label{fig:depth}
\end{figure}

 Generally speaking, the increase of the depth of the neural network is conducive to data fitting, while the increase of the number of hidden layers may also lead to gradient descent or gradient disappearance. In this set of experiments, we increase the number of residual blocks to discuss the influence of depths on performance. When the number of residual block is $i$, we set the number of neurons in each block as $[2^6*2^{i-1},2^6*2^{i-2},\dots,2^6]$. The specific results are recorded in Table \ref{table:depth}, and Figure \ref{fig:depth} shows the performance of several models more visually. 

\begin{table}[htbp]
    \footnotesize
    \centering
    \caption{Performance with respect to the number of residual blocks}
    \subtable[MAE]{
    \setlength{\tabcolsep}{4mm}{
    \begin{tabular}{c|c|c|c|c|c|c}
    \hline
    \textbf{density}&\textbf{5\%}&\textbf{10\%}&\textbf{15\%}&\textbf{20\%}&\textbf{25\%}&\textbf{30\%} \\
    \hline
    \hline
    1 Block &   0.360   &   0.333   &   0.306   &   0.295   &   0.286   &   0.279\\
    2 Block &   0.356   &   0.317   &   0.297   &   0.285   &   0.279   &   0.273\\
    3 Block &   0.350   &   0.316   &   0.298   &   0.289   &   0.274   &   0.271\\
    4 Block &   0.355   &   0.314   &   0.299   &   0.282   &   0.275   &   0.266\\
    \hline 
    \end{tabular}}
    }
    \subtable[RMSE]{
    \setlength{\tabcolsep}{4mm}{
    \begin{tabular}{c|c|c|c|c|c|c}
    \hline
    \textbf{density}&\textbf{5\%}&\textbf{10\%}&\textbf{15\%}&\textbf{20\%}&\textbf{25\%}&\textbf{30\%} \\
    \hline
    \hline
    1 Block &   1.259   &   1.212   &   1.159   &   1.136   &   1.129   &   1.111\\
    2 Block &   1.244   &   1.187   &   1.140   &   1.123   &   1.108   &   1.094\\
    3 Block &   1.243   &   1.182   &   1.136   &   1.114   &   1.097   &   1.088\\
    4 Block &   1.235   &   1.171   &   1.127   &   1.105   &   1.082   &   1.073\\
    \hline 
    \end{tabular}}
    }
    \label{table:depth}
\end{table}
\par It can be seen from the Figure \ref{fig:depth} that under the six densities, the performance is the worst when the number of residual blocks is 1(the number of hidden layers is 2). While MAE performance of the remaining models was similar, the RMSE performance gap was significant. As the number of network layers increases, the RMSE performance of the model also improves. In the baseline approach LDCF, there is almost no performance improvement for more than 4 hidden layers\cite{zhang2019location}, which fully demonstrates that the application of our ResNet greatly reduces the problem of gradient disappearance. This allows PLRes to use deeper networks for better results than existing deep learning methods.

\subsubsection{Impact of Loss Function}\label{subsubsec:Loss_Function}
\par
\begin{figure}[htbp]
    \centering
        \subfigure[MAE]{
            \includegraphics[width=0.45\textwidth]{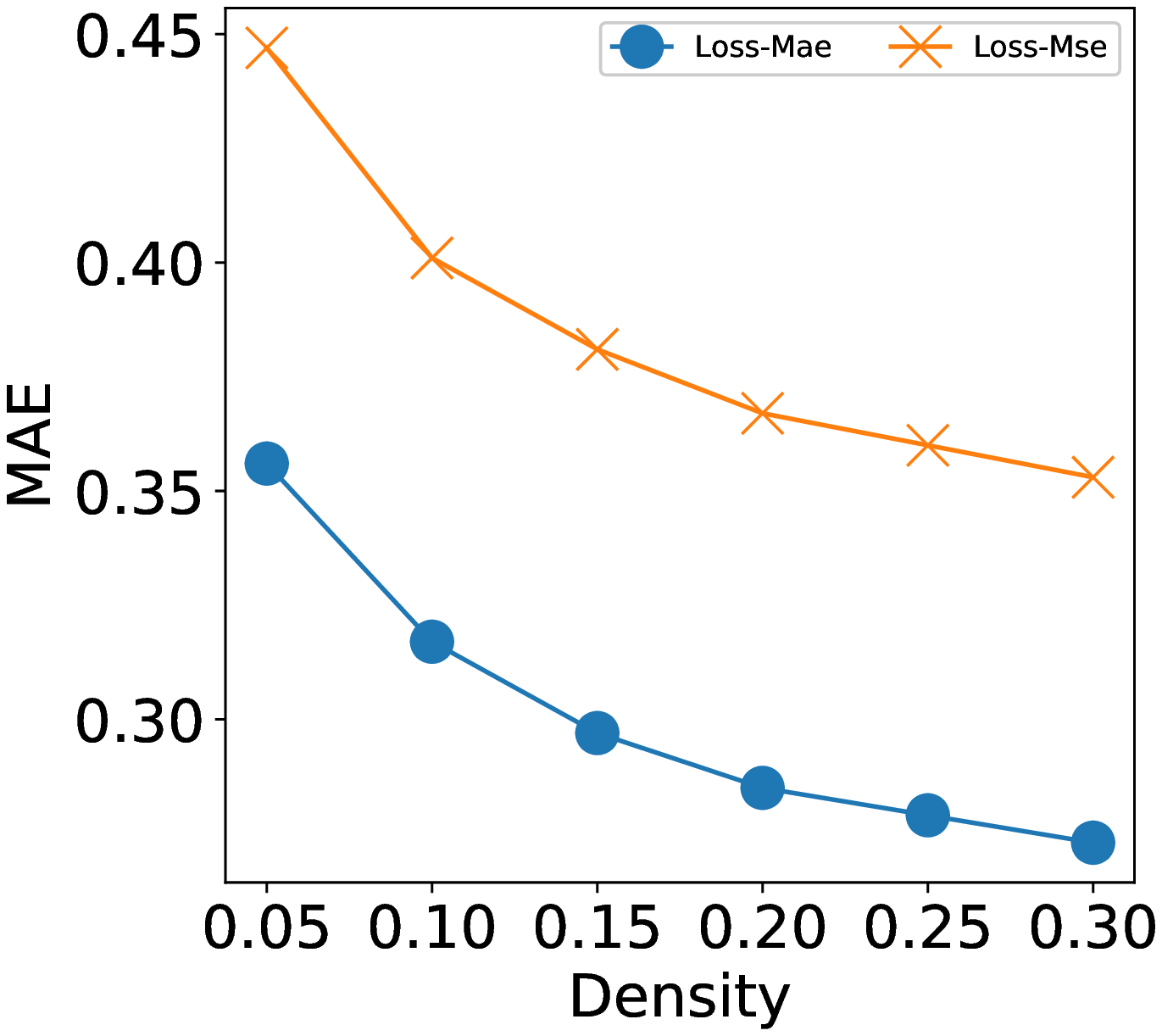}}
        \quad
        \subfigure[RMSE]{
            \includegraphics[width=0.45\textwidth]{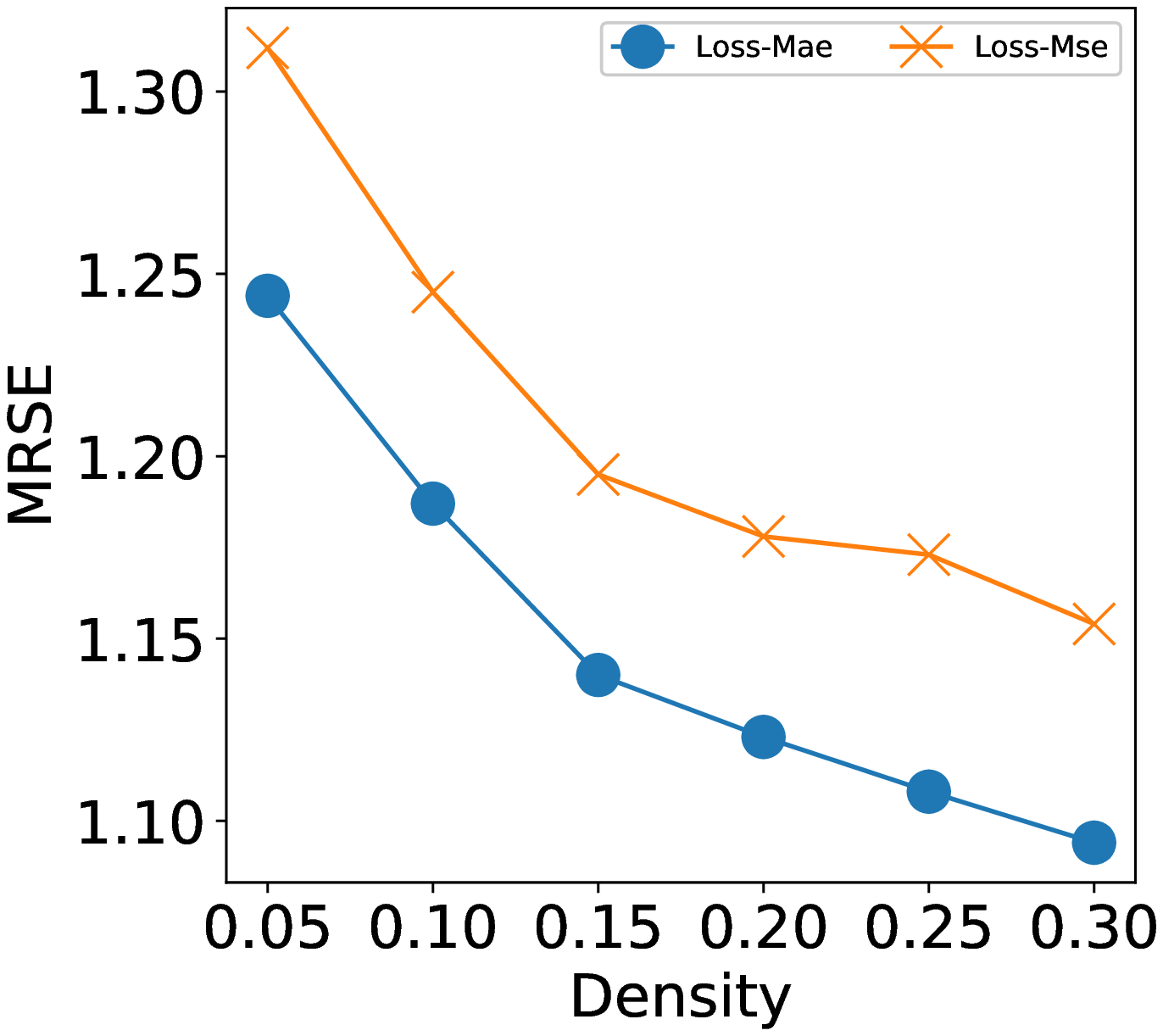}}
    \caption{Impact of loss function.}
    \label{fig:LossFunction}
\end{figure}

 In this set of experiments, we explored the impact of loss functions on the experimental results. According to our performance evaluation method, we used MAE and MSE as loss functions respectively, and use "$Loss$-$Mae$" and "$Loss$-$Mse$" to represent the corresponding models. The number of residual block is set to be 2 and the learning rate is 0.001. In Figure \ref{fig:LossFunction}, we give results with densities of 5\%-30\%. 

It can be seen from the figure that the test results of $Loss$-$Mae$ are much better than those of $Loss$-$Mae$. We choose MAE as the loss function in our model. On the one hand, $Loss$-$Mae$ performs better in both MAE and RMSE on the sparse data; on the other hand, RMSE is greatly affected by outliers, and we pay more attention to the general situation.

\subsubsection{Impact of Learning Rate}
\par
\begin{figure}[htbp]
    \centering
        \subfigure[MAE]{
            \includegraphics[width=0.45\textwidth]{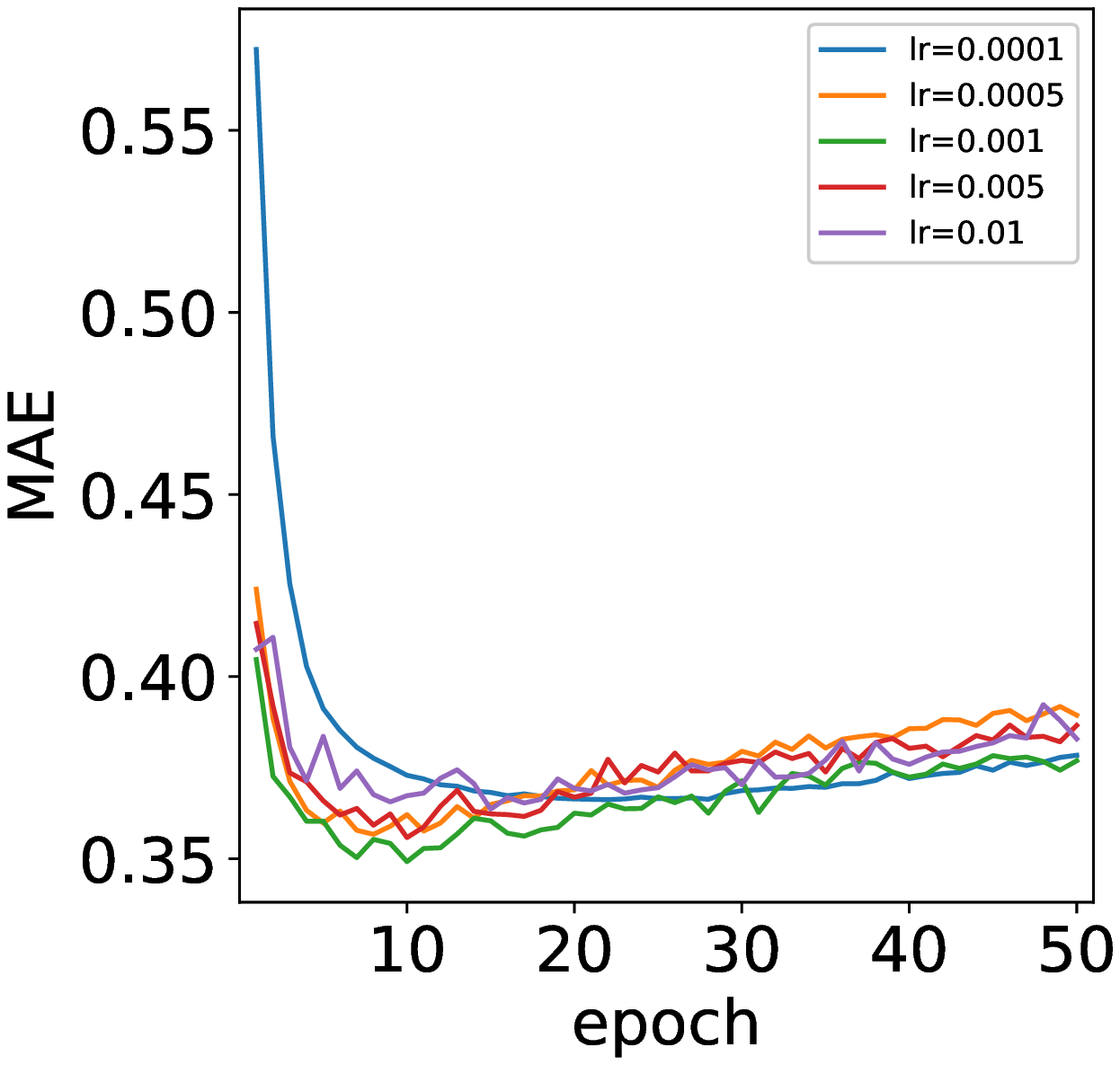}
            \label{lr_mae}}
        \quad
        \subfigure[RMSE]{
            \includegraphics[width=0.45\textwidth]{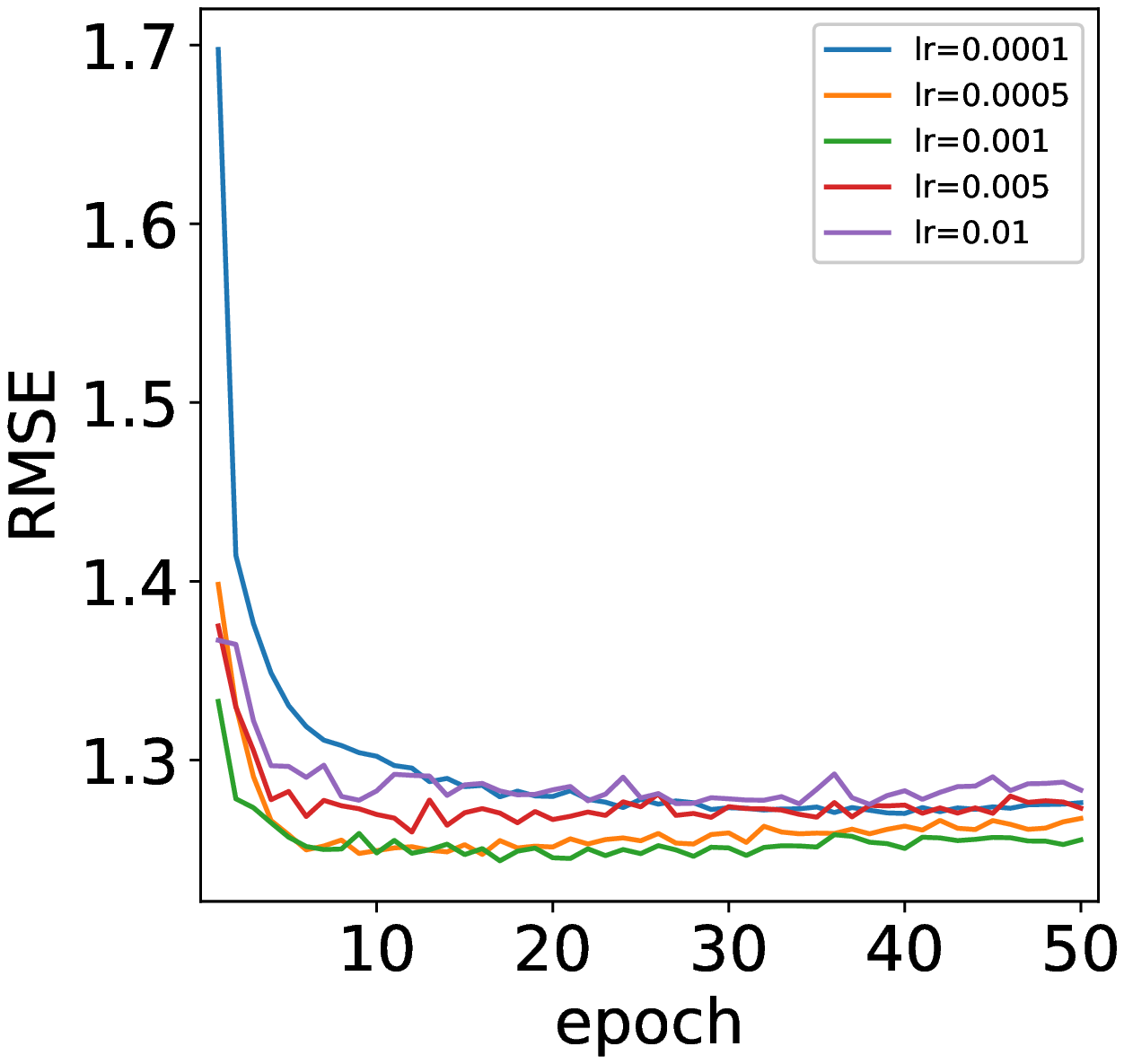}
            \label{lr_rmse}}
    \caption{Impact of learning rate.}
    \label{fig:lr}
\end{figure}

 In the process of model learning, the learning rate affects the speed of model convergence to an optimum. Generally speaking, the higher the learning rate is, the faster the convergence rate will be. While the high learning rate often leads to the failure to reach the optimal solution beyond the extreme value, and a low learning rate often leads to local optimum results. We set up the maximum number of iterations for 50. Figure \ref{fig:lr} shows the change of MAE and RMSE when the learning rate were 0.0001, 0.0005, 0.001, 0.005 and 0.01. 

In the experiment, the models were tested with the testing set when each epoch finish. Therefore, the lowest point of each curve is the optimal result of the corresponding learning rate model, and Table \ref{table:lr} gives the best results of the models under the different learning rates. When the curve in the figure starts to rise, it indicates that the model starts to overfit. In the Figure \ref{fig:lr}, only the curve with a learning rate of 0.0001 is relatively smooth, but its best result is not as good as other models, which is considered the model falls into the local optimal during training. According to the Figure \ref{lr_mae}, which describes the MAE performance, it can be observed that when the epoch reaches 10, the other four curves have reached the lowest point and then started to rise gradually. In terms of RMSE, which is shown in Figure \ref{lr_rmse}, when the epoch reaches 10, the curves gradually tend to be stable. Among these curves, the curve with a learning rate of 0.001 worked best and the curve with the learning rate of 0.0005 is the next. Therefore, when the learning rate is 0.005 and 0.01, we consider the models are difficult to converge due to the high learning rate.
\begin{table}[htbp]
    \footnotesize
    
\setlength{\abovecaptionskip}{0.cm}
 
\setlength{\belowcaptionskip}{0.3cm}
    \centering
    \caption{Experimental results of different learning rate at density 5\%}
    \setlength{\tabcolsep}{5mm}{
    \begin{tabular}{ccc}
    
    \hline
    \textbf{lr}&\textbf{MAE}&\textbf{MRSE}\\
    \hline
    \hline
    0.0001&0.372&1.270\\
    0.0005&0.366&1.247\\
    0.001&0.356&1.244\\
    0.005&0.354&1.260\\
    0.01&0.382&1.275\\
    \hline
    \end{tabular}}
\label{table:lr}
\end{table}

\section{Discussion}\label{sec:Discussion} 
\noindent
\par In this section, we discuss why PLRes works. Specifically, why the use of probability distribution, location information, and the introduction of ResNet help improve model performance.

\subsection{The Advantages of Using the Probability Distribution}
Probability distribution is the probability of missing QoS value in each interval, which is represented by the distribution of historical invocations of users and services in our approach. This distribution is an intuitive representation of the historical data. The historical data is the basis of prediction, so the use of probability distribution as a feature is beneficial to the prediction of Qos in theory.
\begin{figure}[htbp]
    \centering
        \subfigure[Distribution of service invocations for 5 random users]{
            \includegraphics[width=1.0\textwidth]{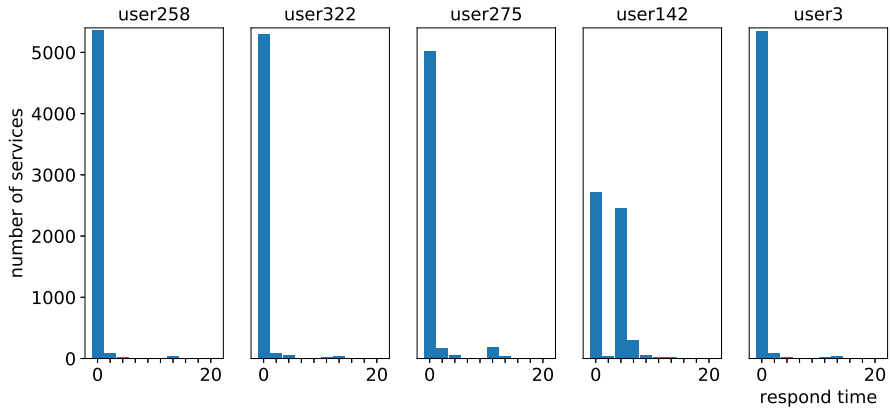}
            \label{user_pro}}
        \subfigure[Distribution of service invocations for 5 random services.]{
            \includegraphics[width=1.0\textwidth]{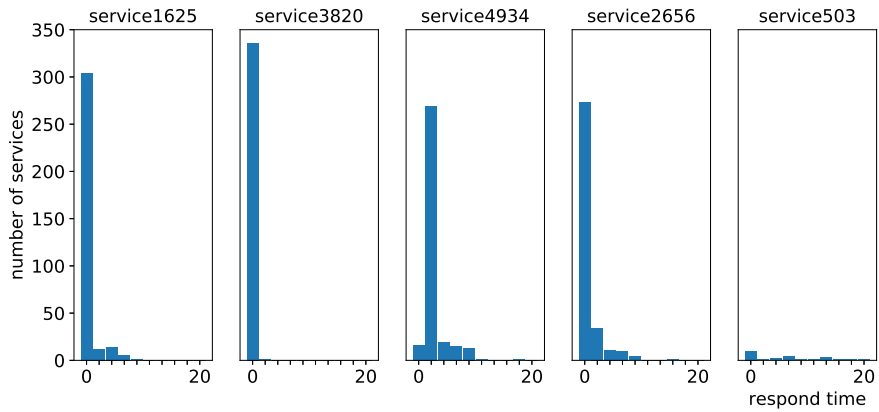}
            \label{service_pro}}
    \caption{Distribution of service invocations random samples.}
    \label{fig:probability distribution}
\end{figure}
\par We randomly selected samples from the original dataset and calculated their distribution of service response time. Figure \ref{user_pro} shows the distribution of historical invocation for five different users and Figure \ref{service_pro} is the services. The abscissa axis represents the time interval and the ordinate axis represents the number of services invoked by the user. Take the $user258$ in Figure \ref{user_pro} as an example, there are 5366 services whose response time is less than $2s$, 92 services whose response time is greater than $2s$ and less than $4s$, and so on until $20s$. As can be seen from the Figure \ref{user_pro}, the service response time distribution of several users is mainly concentrated within $2s$, but the distribution of $user142$ is quite different. In fact, we also randomly checked the QoS distribution of some other users, most of which were similar to $user258$ and $user322$, while the distribution of a small number of users was quite different from that of other users. The historical distribution of services also shows a similar pattern: the response times of most services are similar to those of $service1625$, $service3820$, and $service2656$ in Figure \ref{service_pro}, while a few services are abnormal, such as $service4934$ and $service503$. Therefore, probability distribution is helpful in reflecting sample characteristics. Considering the total number of historical invocations by different users or services are always different, we do not use the distribution of the number of invocations, but the proportion of distribution as a feature.
\par The probability distribution not only reflects the user preference but also effectively reflects the network stability of users and services. Therefore, the introduction of probability distribution is helpful to reduce the sensitivity to abnormal data and to reduce the overfitting of the model.

\subsection{The Advantages of Using Location Information}
\par According to the results in Section \ref{subsec:effect_loc}, The analysis of the location characteristic is extremely advantageous for predicting QoS value. Users in the same region often have similar network conditions, while the network status of users in different areas are usually differ greatly. Therefore, location information can be used as an important reference factor for user similarity, which is also the reason why partial collaborative filtering methods use location information. In addition, location information can reflect the distance between the user and the server, which also tends to affect service efficiency. Even if the invoked service is the same one, users who closer to the server always get better network response and bandwidth. 
What's more, the introduction of location characteristic is also helpful to solve the cold start problem. Even the users who have never invoked any service can give valuable predictions of QoS based on their geographic position and the historical invocations of the same location learned by the model.
\par Another advantage of location information is that it is more accessible than more complex features, which could be easily queried by the IP address of the user or service. The validity of location information for QoS prediction can also be known from a great deal of literature.

\subsection{The Advantages of Using ResNet}
\par In the prediction of QoS, CF is the most commonly used, but the limitations of CF are also apparent. Using the model-based collaborative filtering method MF, the latent vectors of users and services can be obtained, but the inner product function can limit the expressiveness of MF in the low-dimensional latent space.\cite{DBLP:conf/www/HeLZNHC17} Therefore, in the latent space, the similarity between users (or services) may be quite different from the result obtained by Jaccard similarity, incurring a large loss of similarity ranking\cite{zhang2019location}.

\par
\begin{figure}[htbp]
\centerline{\includegraphics[width=0.4\textwidth]{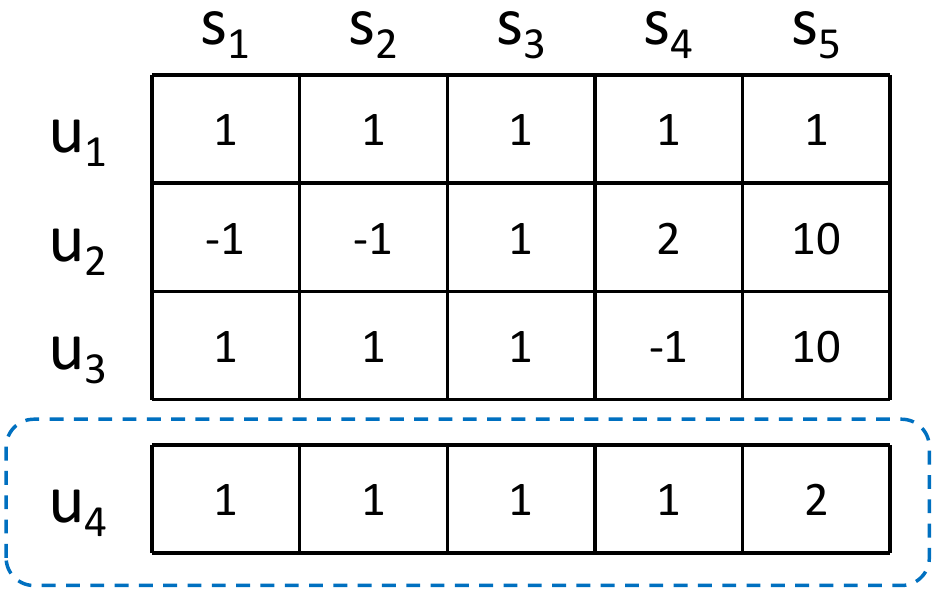}}
\caption{Example of PCC similarity calculation.}
\label{example}
\end{figure}

 In memory-based collaborative filtering, Pearson Correlation Coefficient(PCC) is a common method to calculate similarity. The calculation method is as equation (\ref{PCC}):
\begin{equation}
Sim_{PCC}(u,v)=\frac{\Sigma_{i\in I}(q_{u,i}-\bar{q}_{u})(q_{v,i}-\bar{q}_{v})}{\sqrt{\Sigma_{i\in I}(q_{u,i}-\bar{q}_{u})^{2}}\sqrt{\Sigma_{i\in I}(q_{v,i}-\bar{q}_{v})^{2}}}\label{PCC}
\end{equation}
where $I$ represents the intersection of the services invoked by user $u$ and $v$, $q_{u,i}$ and $q_{v,i}$ represent the QoS value of $u$ and $v$ invoking service $i$ respectively, $\bar{q}_{u}$ and $\bar{q}_{v}$ denote the average QoS value of $u$ and $v$ invoking services respectively. But in some cases, this method is not very effective. For example, Figure \ref{example} shows the response times for four users to invoke five services, $-1$ represents no invoke record. It could be easily observed that the user $u_1$ is the most similar to the user $u_4$ among the first three users. They invoked the same services and had similar response times. However, according to the calculation of PCC, it can be calculated that Sim($u_{1}$,$u_{4}$) = 0 $<$ Sim($u_{2}$,$u_{4}$) = 0.994 $<$ Sim($u_{3}$,$u_{4}$) = 1. Therefore, the use of deep learning could avoid this kind of similarity calculation and such similarity errors.

\par In deep learning networks, the increase of network layers is conducive to learning higher dimensions features and solving more complex tasks. However, the reality is that the increase of network depth results in feature loss and gradient disappearance. This is also the reason why the performance improvement of deep learning networks for QoS prediction is not obvious even if the depth increases. The ResNet\cite{DBLP:conf/cvpr/HeZRS16} is proposed in the field of image recognition to solve the problem of gradient disappearance. It effectively alleviates the problem by using shortcuts to enhance the characteristics in the process of model training. Although the shortcuts of the ResNet mainly connect the convolutional layers, the idea of feature reuse is also applicable in the full connection layer. In QoS prediction, the features we can collect are limited and brief, so it is a good choice to use ResNet to capture high-dimensional nonlinear features and enhance them during model training. 


\section{Related Work}\label{sec:RelatedWork}  
\noindent
\par In the existing QoS prediction methods, collaborative filtering is the most widely used technology. Collaborative filtering fully considers the user's preference, so it is commonly used in the recommendation system and performs well in the personalized recommendation.
\par Collaborative filtering methods can be divided into two categories: memory-based and model-based. The memory-based collaborative filtering method usually achieves the prediction of the target by obtaining similar users or services with similar neighbors. Therefore, memory-based collaborative filtering can be subdivided into user-based, service-based and hybrid-based. Linden et.al\cite{DBLP:journals/internet/LindenSY03} help the recommend system find similar items of what the user needs and add them to the recommended sequence by the item-to-item collaborative filtering. Adeniyi et.al\cite{ADENIYI201690} used K-Nearest-Neighbor (KNN) classification method to find similar items for recommendation systems. Zou et.al\cite{DBLP:conf/icsoc/ZouJNWPG18} improved the method to integrate similar users and services, proposed a reinforced collaborative filtering approach. In the model-based collaborative filtering, machine learning method is used to study the training data to achieve the prediction of QoS. Matrix factorization is the most typical and commonly used model-based method, which turns the original sparse matrix into the product of two or more low-dimensional matrices. In QoS prediction, matrix factorization often captures the implicit expression of users and services. Zhu et.al\cite{DBLP:journals/tpds/ZhuHZL17} propose an adaptive matrix factorization approach to perform online QoS prediction. Wu et.al\cite{DBLP:conf/icsoc/WuXCCZ17} using the FM(Factorization Machine approach), which is based on MF to predict missing QoS values. Tang et.al\cite{DBLP:journals/access/TangLYX19} considered the similarity as a character, proposed a collaborative filtering approach to predict the QoS based on factorization machines. However, CF can only learn linear features, so many methods begin to consider in-depth learning that can effectively learn nonlinear features.

\par Deep learning is a subset of machine learning, and it combines the characteristics of the underlying data to form a more abstract and deep representation. Due to its strong learning ability on hidden features, it has been widely used in various recommendation systems\cite{10.1145/3285029,9017973,9004579,WANG202068}.
\par In QoS prediction, some methods combine deep learning with collaborative filtering. Zhang et.al\cite{zhang2019location} proposed a new deep CF model for service recommendation to captures the high-dimensional and nonlinear characteristics. Soumi et.al\cite{DBLP:conf/icsoc/ChattopadhyayB19} proposed a method which is a combination of the collaborative filtering and neural network-based regression model. Xiong et.al\cite{xiong2018deep} proposed a deep hybrid collaborative filtering approach for service recommendation (DHSR), which can capture the complex invocation relations between mashups and services in Web service recommendation by using a multilayer perceptron. Deep learning is also often used in methods using the timeslices of service invocation. Xiong et.al\cite{DBLP:conf/icws/XiongWLLH18} propose a novel personalized LSTM based matrix factorization approach that could capture the dynamic latent representations of multiple users and services. Hamza et.al\cite{DBLP:conf/icsoc/LabbaciMA17} uses deep recurrent Long Short Term Memories (LSTMs) to forecast future QoS.

\par In some existing researches\cite{8029908, DBLP:journals/mis/LiWW19, 7799646, DBLP:conf/icws/LoYDLW12,DBLP:journals/ijcse/0007SYML19}, location information is considered as one of the characteristics of QoS prediction. Li et.al\cite{li2017location} propose a QoS prediction approach combining the user's reputation and geographical information into the matrix factorization model. Tang et.al\cite{doi:10.1002/cpe.3639} exploits the users' and services' locations and CF to make QoS predictions. Chen et.al\cite{DBLP:journals/ijcse/0007SYML19} propose a matrix factorization model that using both geographical distance and rating similarity to cluster neighbors. These approaches have improved the accuracy of QoS prediction, and their experimental results fully demonstrate the validity of location information.


\section{Conclusion and Future Work}\label{sec:Conclusion}
\noindent
\par In this paper, we propose a probability distribution and location-aware QoS approach based on ResNet named PLRes. The model uses ID, location information and probability distribution as the input characteristics. PLRes encodes the ID and geographic location of the users and services, and embedded them into the high-dimensional space. Then all the features(the embedded ID and location features, and the probability distribution) are concatenated into a one-dimensional vector and input into ResNet for learning. We trained the model and conducted experiments on the WS-DREAM dataset. The experimental results fully prove that the features of location and probability distribution are conducive to improving the accuracy of the QoS prediction model. As a deep learning method, PLRes gives full play to its advantages in learning high-dimensional nonlinear characteristics, and compared with the advanced deep learning method LDCF, PLRes effectively alleviates the gradient disappearance problem.
\par In the future, we will further consider the combination of the current model and collaborative filtering method to make full use of the advantages of collaborative filtering. In addition, we did not consider the time factor for the user to invoke the service in this paper. Since the service is constantly updated, the QoS of different timeslices may change greatly, so the time feature is also necessary in QoS prediction. We will further consider predicting missing QoS value through QoS changes of different time slices in the next work.

\section*{Acknowledgements}
The work described in this paper was partially supported by the NationalKey Research and Development Project (Grant no.2018YFB2101201), the National Natural Science Foundation of China (Grant no.61602504), the Fundamental Research Funds for the Central Universities (Grant no. 2019CDYGYB014).




\bibliography{refs}

\end{document}